\documentclass[smallextended]{svjour3}

\smartqed 
\usepackage{graphicx,amssymb,amsmath,color}
\usepackage{hyperref}

\def \d {{\rm d}}

\begin{document}

\title{Ultrarelativistic boost with scalar field}

\author{O. Sv\'{\i}tek
        \and
        T. Tahamtan
}


\institute{Institute of Theoretical Physics, Faculty of Mathematics and Physics,
Charles University in Prague, V Hole\v{s}ovi\v{c}k\'{a}ch 2, 180~00 Prague
8, Czech Republic\\
\email{ota@matfyz.cz}
\and
\email{tahamtan@utf.mff.cuni.cz}
}

\date{Received: date / Accepted: date}

\maketitle

\begin{abstract}
We present the ultrarelativistic boost of the general global monopole solution which is parametrized by mass and deficit solid angle. The problem is addressed from two different perspectives. In the first one the primary object for performing the boost is the metric tensor while in the second one the energy momentum tensor is used. Since the solution is sourced by a triplet of scalar fields that effectively vanish in the boosting limit we investigate the behavior of a scalar field in a simpler setup. Namely, we perform the boosting study of the spherically symmetric solution with a free scalar field given by Janis, Newman and Winicour. The scalar field is again vanishing in the limit pointing to a broader pattern of scalar field behaviour during an ultrarelativistic boost in highly symmetric situations.
\end{abstract}

\section{Introduction}
It is well known that different types of topological objects may have formed during the initial phase of the expansion of Universe, the most interesting ones being domain walls, cosmic strings and monopoles \cite{Kibble}. These topological defects are assumed to be created during spontaneous breakdown of local or global gauge symmetries. The main feature of the global monopole \cite{Barriola} is the presence of a solid deficit angle which is sourced by a triplet of scalar fields. This solution has enjoyed continuous interest in the past 25 years. It was generalized to both positive and negative cosmological constant \cite{LiHao,Bertrand} and modified theories of relativity \cite{Teixeira,Lee}. Its classical stability was proven for a large class of perturbations \cite{Achucarro}. Since it is a potentially important astrophysical source providing information about the initial phases of our universe its gravitational lensing features were investigated \cite{Perlick} and the scaling properties of networks of global monopoles were analyzed \cite{Martins}. The global monopole solution was also used to study the possible removal of a classical singularity from the quantum mechanical point of view --- in the case of a Klein-Gordon field \cite{Pitelli}, Dirac field and from the Loop Quantum Gravity perspective \cite{Tahamtan}.

Ultrarelativistic limits of solutions to the Einstein equations generally transform stationary spacetimes into plane wave solutions with a profile function corresponding to a shock wave. The first such limit was performed by Aichelburg and Sexl \cite{Aichelburg} (for a generalization to sources with higher multipole moments, see \cite{Podolsky}). Their procedure of boosting the spacetime to the speed of light consists of taking a proper distributional limit accompanied by a specific coordinate transformation that provides a regularization of the result. This fact brings a certain ambiguity which is similar to the one associated with the renormalization in Quantum Field Theory as noted by \cite{Balasin1}. This led the authors of \cite{Balasin1} to propose an alternative route by performing an ultrarelativistic limit of energy momentum tensor of the given spacetime instead of the metric. This meant finding a proper distributional source for the given geometry---for Schwarzschild it is unsurprisingly a density distribution in the form of Dirac's $\delta$-function located at the origin. The geometry is then recovered by solving Einstein equations for the ultraboost limit of distributional energy momentum tensor.

The ultrarelativistic boost of a global monopole was previously derived in \cite{Lousto} but our result is different. We will explain where the difference arises and support our conclusion by comparing the results coming from the above mentioned approaches to boosting.

Since the triplet of scalar fields and the associated energy momentum tensor is shown to vanish in the ultraboost limit we try to investigate the behaviour of the scalar field during boosting for a simpler case of a single scalar field with spherical symmetry --- namely the so-called Janis--Newman--Winicour (JNW) spacetime. The aim is to confirm that the scalar fields in these highly symmetric situations generally tend to vanish and that the result obtained for the global monopole is not surprising.

In 1968 Janis, Newman and Winicour \cite{JNW} presented the most general spherically symmetric, static and as\-ymptotically flat solution of Einstein's field equation minimally coupled to a massless scalar field. They main\-ly discussed its specific "vacuum" limit. After dealing with a certain ambiguity in the limiting process they discovered that their solution tends to the exterior of Schwarz\-schild but with a pointlike singularity at the horizon position. Later, this solution was rediscovered in a different form by Wyman \cite{Wyman} (see \cite{Virbhadra} for an explanation of the coincidence of both solutions). In fact, although the name of the spacetime is widely used in the literature, a recent paper \cite{Tafel} shows that it is just one in the series of rediscoveries of the solution originally published by Fisher \cite{Fisher}.

The peculiar behaviour of the JNW solution with respect to the Schwarzschild limit at the horizon position was generalized in \cite{Tafel} to cover all spherically symmetric, static and asymptotically flat (or anti-de Sitter) solutions of Einstein's field equations minimally coupled to a scalar field with a potential. Specifically, it states that the global structure of these geometries is Schwarzschild(--anti-de Sitter)-like with naked singularity potentially shifted to a nonzero radius.

In this paper, we first apply the two boosting methods mentioned above to the general metric of a global monopole \cite{Barriola}, which is determined by two parameters --- one characterizing the "Schwarzschild mass" and the other measuring the deficit solid angle. More precisely, we use a line element which is valid only outside the core of the global monopole source (see e.g. \cite{Bezzera} for an extensive discussion) but can as well be interpreted as the exact metric corresponding to a spherically symmetric cloud of strings \cite{Letelier}. Since even in the context of the cloud of strings the metric is referred to as the global monopole solution we keep using this name while having both interpretations (and their limitations) in mind. The considered solution possesses a singularity which is generally covered by a single horizon but this two parameter class of metrics also contains, as a special case, a naked singularity. Finally, we will review the JNW metric and subsequently perform its boost to the velocity of light again using both approaches. In both cases we discuss the resulting geometries and analyze the consequences.

\section{Global Monopole}

The simplest scalar field model with a minimal coupling (see \cite{Nucamendi} for a generalization to non-minimal couplings) giving rise to the global monopole is described by the Lagrangian \cite{Barriola}
\begin{equation}
L=\frac{1}{2}\partial _{\mu }\phi ^{a}\partial ^{\mu }\phi ^{a}-\frac{1}{4}\lambda (\phi ^{a}\phi ^{a}-\eta^{2})^{2},
\end{equation}
where $\phi ^{a}$ is a triplet of scalar fields, $a=1,2,3$ and $\eta, \lambda$ are constants. The model has a global $O\left( 3\right) $ symmetry, which is spontaneously broken to $U\left( 1\right)$. The scalar field triplet corresponding to a monopole has the following form
\begin{equation}
\phi ^{a}=\eta \frac{x^{a}}{r}f(r)
\end{equation}%
where $x^{a}x^{a}=r^{2}$. The function $f\sim 1$ outside the monopole core which has the size $\sim\lambda^{\frac{-1}{2}}\eta^{-1}$ and we will consider a spacetime given by $f=1$ from now on. 

We assume that the underlying geometry is described by a general static spherically symmetric line element
\begin{equation}\label{general-metric}
ds^{2}=-B(r) dt^{2}+\frac{dr^{2}}{A(r) }+r^{2}\left(
d\theta ^{2}+\sin ^{2}\theta d\phi ^{2}\right) ,
\end{equation}%
with a standard relation between spherical, $\{r,\theta ,\phi\}$, and cartesian-like coordinates $x^{a}$. The Lagrangian for the above given
field configuration and geometry takes this form%
\begin{equation}
L=\frac{1}{2}\left( \partial _{\theta }\phi ^{a}\partial ^{\theta }\phi
^{a}+\partial _{\phi }\phi ^{a}\partial ^{\phi }\phi ^{a}\right) =\frac{\eta
^{2}}{r^{2}},
\end{equation}%
giving the following energy momentum tensor %
\begin{equation}\label{EMT}
T_{t}^{t}=T_{r}^{r}=-\frac{\eta ^{2}}{r^{2}},\text{ \ \ }T_{\theta }^{\theta
}=T_{\phi }^{\phi }=0.\text{\ }
\end{equation}%
Solution of the Einstein equations for metric (\ref{general-metric}) with energy momentum tensor (\ref{EMT}) is described by the following functions
\begin{equation}\label{AB-functions}
B(r)=A(r)=1-8\pi G\eta ^{2}-\frac{2GM}{r}
\end{equation}%
where $M$ is a constant of integration. This metric describes a black hole of mass $M$, carrying a global mono\-pole charge $\eta$ which measures the deficit of solid angle. One can imagine this object being formed when an ordinary black hole swallows a global monopole \cite{Kibble}. As already mentioned in the introduction, the above interpretation of metric (\ref{general-metric}) with (\ref{AB-functions}) is valid only outside the core of the source located at the origin. An alternative explanation of this line element comes from considering a spherically symmetric cloud of strings (generalization of a dust cloud) \cite{Letelier} with energy density equal to tension \cite{Pitelli}.

The Kretschmann scalar which indicates the presence of a curvature singularity is given by%
\begin{equation}
\mathcal{K}=\frac{48M^{2}G^{2}}{r^{6}}+\frac{128M\pi G^{2}\eta ^{2}}{r^{5}}+%
\frac{256\pi ^{2}G^{2}\eta ^{4}}{r^{4}}.
\end{equation}%
Obviously, at $r=0$ we have a typical central curvature singularity and the dominant contribution comes from the Schwarzschild mass.

\section{Kerr-Schild form of metric}\label{III}

To arrive at the Kerr-Schild form of metric \cite{KerrSchild} suitable for both approaches to boosting we need to perform certain simple coordinate transformations. Let us first transform the metric to isotropic coordinates. Starting from metric (\ref{general-metric}) with (\ref{AB-functions}) we can write
\begin{equation}
ds^{2}=A(r) \left(-dt^{2}+\frac{dr^{2}}{A(r)^{2} }\right)+r^{2}\left(
d\theta ^{2}+\sin ^{2}\theta d\phi ^{2}\right)
\end{equation}
and define $d\mu=\frac{1}{A(r)} dr-dt$. The line element will become 
\begin{eqnarray}
ds^{2}&=&-A(r)d\mu^{2}+2d\mu\,dr+r^{2}d\Omega^{2}=\\
&=&-d\mu^{2}+2d\mu\,dr+r^{2}d\Omega^{2}+[1-A(r)]d\mu^{2}\nonumber
\end{eqnarray}
By changing the coordinates again using $d\mu=dr-dt'$ we get the final form of the metric 
\begin{equation}\label{KS-form}
ds^{2}=-dt'^{2}+dr^{2}+r^{2}d\Omega^{2}+\left[1-A(r)\right](dr-dt')^{2},
\end{equation}
in which $A(r)=1-8\pi G\eta ^{2}-\frac{2GM}{r}$. We will rewrite the metric in corresponding cartesian-like coordinates (which reflect the flat part of the Kerr-Schild form of metric that one can treat as a background)
\begin{eqnarray}\label{Cartesian}
ds^{2}&=&-dt'^{2}+dx^{2}+dy^{2}+dz^{2}+\\
&&+\frac{1-A(r)}{r^{2}}\left(xdx+ydy+zdz-rdt'\right)^{2},\nonumber
\end{eqnarray}
where $r^{2}=x^{2}+y^{2}+z^{2}$. So the metric is separated into a Minkowski part and one additional term encoding the Schwarzschild black hole with a global monopole.

\section{Boosting global monopole}
The Minkowski part of the Kerr-Schild form of metric provides us with a notion of boosts as isometries of this background \cite{Balasin1} thus acting similarly to gauge transformations (e.g. in gravitational waves on a background) in this respect. Additionally, an asymptotic observer moving uniformly relative to the origin of this spherically symmetric spacetime will see the metric deformed by a Lorentz transformation \cite{Aichelburg} if we have only $M$ nonvanishing. We choose a Lorentz transformation in the $x-$direction
\begin{equation}\label{Lorentz}
\bar{t}=\frac{(t'+vx)}{\sqrt{1-v^{2}}},\quad \bar{y}=y,\quad \bar{z}=z,\quad \bar{x}=\frac{(x+vt')}{\sqrt[]{1-v^{2}}},
\end{equation}
where $v$ is the boost parameter. The Minkowski part of the line element (\ref{Cartesian}) is obviously invariant under Lorentz transformation so we only study the additional term that is modified as follows
\begin{eqnarray}\label{transformedChi}
\chi(\bar{r})[(\bar{x}-v\bar{t})(d\bar{x}-vd\bar{t})+(1-v^{2})(ydy+zdz)-&&\nonumber\\
-\sqrt{(\bar{x}-v\bar{t})^{2}+(1-v^{2})(y^{2}+z^{2})}(d\bar{t}-vd\bar{x})]^{2},&&
\end{eqnarray}
where
\[\chi(\bar{r})=\frac{8\pi G\eta ^{2}\bar{r}+2GM}{\bar{r}^{3}(1-v^{2})^{2}}=\]
\begin{equation}
\frac{2pG(1-v^{2})+8\pi G\eta ^{2}\sqrt{(\bar{x}-v\bar{t})^{2}+(1-v^{2})(y^{2}+z^{2})}}{(1-v^{2})[(\bar{x}-v\bar{t})^{2}+(1-v^{2})(y^{2}+z^{2})]^{3/2}}
\end{equation}
Here we automatically consider the same rescaling of the mass as in the original paper \cite{Aichelburg}
\begin{equation}\label{scaling}
	M=p\sqrt{1-v^{2}},
\end{equation}
with $p$ being a constant throughout the limiting process. 

Now, we will study the limit of (\ref{transformedChi}) when $v\rightarrow 1$. The limit separates into two cases in the following way
\begin{equation}
\begin{cases}
\frac{32\pi G\eta ^{2}|\bar{x}-\bar{t}|+8GM{\sqrt[]{1-v^{2}}}}{|\bar{x}-\bar{t}|({1-v^{2}} )}(d\bar{x}-d\bar{t})^{2}  & \bar{x} \neq \bar{t}\\
{}\\
\frac{8\pi G\eta ^{2}\rho+2GM}{\rho({1-v^{2}} )}(d\bar{x}-d\bar{t})^{2} &  \bar{x} = \bar{t}
\end{cases}
\end{equation}
in which ${\rho}^2=(y^{2}+z^{2})$. It is clear that for a nonvanishing $\eta$ the two limits are not finite and cannot be regularized in the manner of \cite{Aichelburg}. It is interesting to note that if $M=0$, the limits coincide for both cases (up to a numerical factor). However, generally it is possible to rescale $\eta$ in the same way as we did for mass, i.e. 
\begin{equation}\label{angle-scaling}
	\eta=\eta_{0}\sqrt{1-v^{2}},
\end{equation}
then the limit becomes trivial
\begin{equation}
\begin{cases}
\left[ 32\pi G\eta_{0} ^{2}+\frac{ 8Gp}{|\bar{x}-\bar{t}|}\right](d\bar{x}-d\bar{t})^{2}  & \bar{x} \neq \bar{t}\\
{}\\
\left[8\pi G\eta_{0} ^{2}+\frac {2GMp}{\rho \sqrt[]{1-v^{2}}}\right](d\bar{x}-d\bar{t})^{2} &  \bar{x} = \bar{t}
\end{cases}
\end{equation}
When $\eta=0$, we will retrieve the Schwarzschild case which is finite when $\bar{x} \neq \bar{t}$ and leads to a flat space. In the case of $\bar{x}=\bar{t}$ one can perform the same regularization trick (first a singular transformation and then a boost parameter limit) that one can find in \cite{Aichelburg} and it ensures the resulting metric might be expressed in terms of Dirac's delta distribution (multiplied by $\ln{\rho}$) corresponding to a shock wave located on the $\bar{x}=\bar{t}$ null surface. Alternatively, one can first perform a distributional limit of metric functions using the Hotta and Tanaka identity \cite{Hotta}
\begin{equation}\label{distributional-limit}
	\lim_{v\to 1}\frac{1}{\sqrt{1-v^{2}}}f(\bar{x})=\delta(x-t)\int_{-\infty}^{\infty}{f(\xi)}\d \xi
\end{equation}
and then regularize the infinities resulting from the unbounded integral (that are of the form $\ln(\epsilon)$ for $\epsilon\to \infty$ in the case of Schwarzschild) using renormalization techniques from Quantum Field Theory.

A different solution to the problem of ultrarelativistic limit when $\eta\neq 0$ was proposed in \cite{Lousto} where the suggested scaling for the monopole charge is $\eta=\eta_{0}(1-v^{2})^{1/4}$. In that case, however, the limits still remain infinite both inside and outside the null hyperplane $\bar{x}=\bar{t}$. On the other hand, performing the limit according to (\ref{distributional-limit}) leads to a shock wave profile $~\delta(x-t){\rho}$. But here one needs to remove the terms linear in $\epsilon$ (for $\epsilon\to \infty$) by a suitable regularization in all spacetime points.

From the physical perspective one can also study the behavior of volume during boosting since it is directly related to the interpretation of $\eta$ as a parameter describing the deficit of solid angle. For this purpose we shall consider the essential part of the spatial volume three-form ${}^{3}\omega=\sqrt{{}^{3}g}\,\d x\wedge\d y\wedge\d z$ for the spatial metric ${}^{3}g_{ij}$ (with ${}^{3}g=\det{{}^{3}g_{ij}}$) induced from (\ref{Cartesian}), namely the scalar density
\begin{equation}
\sqrt{{}^{3}g}=\sqrt{2-A(r)}=\sqrt{1+8\pi G\eta ^{2}+\frac{2GM}{r}}\ .
\end{equation} 
Concentrating solely on the influence of $\eta$ (by putting $M=0$) we see that in the ultraboost limit both scalings lead to a vanishing contribution of $\eta$ to the volume. Now, let us consider the behavior of the volume element which gains an additional $\gamma$-factor when boosted and set general scaling $\eta=\eta_{0}(1-v^2)^{s}$
\begin{equation}
{}^{3}\omega \sim \gamma\sqrt{1+8\pi G\eta_{0}^{2}(1-v^2)^{2s}}=\sqrt{\frac{1}{1-v^2}+8\pi G\eta_{0}^{2}(1-v^2)^{2s-1}}\ .
\end{equation}
For $s=1/2$ one can interpret $\eta_{0}$ as giving fixed contribution to volume during boosting process. This value coresponds to proposed scaling (\ref{angle-scaling}).

Note that the scalar field triplet, its Lagrangian and energy momentum tensor are all going to zero for both scalings. In the next section we present alternative approach to the problem to help us with deciding which scaling should be preferred.

\section{Regularized energy momentum tensor limit}

Different approach to boosting provides a method developed by Balasin and Nachbagauer \cite{Balasin1}. It is based on the analysis of the energy momentum tensor as the primary object describing the spacetime rather than the metric. Note that the energy momentum tensor is considered here to be defined by the left-hand side of the Einstein equations. Since the Schwarzschild solution is without a source in the standard sense this would not reproduce the previous results \cite{Aichelburg}. First, one has to define a correct energy momentum tensor of a given solution in terms of distributions \cite{Balasin2} (making the connection between the gravitational and the electrostatic field of a particle even more explicit). Next, one performs an ultrarelativistic boost of such an energy momentum tensor and takes the result as a source for a new spacetime corresponding to the boosted geometry.

Let us review the construction of the distributional energy momentum tensor for the Kerr-Schild class of spacetimes as introduced in \cite{Balasin2}. The metric for such a solution can be given in the following form
\begin{equation}
	g_{ab}=\eta_{ab}+fk_{a}k_{b} ,
\end{equation} 
where $\eta_{ab}$ is the Minkowski metric, $f$ is a function and $k_{a}$ is a covector which is null with respect to both metrics. The metric of a global monopole can be given in this form as explicitly derived in Section \ref{III}. Importantly, the covector field $k_{a}$ satisfies the field equation $G^{ab}k_{a}k_{b}=0$ even in the presence of the scalar field supporting the global monopole thus maintaining the geodetic property discovered by Kerr and Schild \cite{KerrSchild}. Using metric $\eta_{ab}$ for raising/lowering indices and its associated derivative $\partial_{a}$ the Ricci tensor and scalar can be expressed in the following way
\begin{equation}\label{Ricci}
	R^{a}_{b}=\frac{1}{2}(\partial^{a}\partial_{c}(fk^{c}k_{b}) + \partial_{b}\partial^{c}(fk_{c}k^{a}) - \partial^{2}(fk^{a}k_{b})),
\end{equation} 
\begin{equation}\label{Scalar}
	R=\partial_{a}\partial_{b}(fk^{a}k^{b}).
\end{equation}
Both expressions are linear in $f$ which is useful in combining effects of different terms appearing in $f$. 

The distributional evaluation of these expressions then proceeds in the following way \cite{Balasin1} (we give an illustration in the case of a scalar curvature and a test function $\phi\in C^{\infty}_{0}(\mathbb{R}^{3})$, due to the stationarity of the solution we consider only the spatial distributions)
\begin{eqnarray}
	(R,\phi)&=&(\partial_{a}\partial_{b}(fk^{a}k^{b}),\phi)=(fk^{a}k^{b},\partial_{a}\partial_{b}\phi)=\nonumber\\
	&=&\lim_{\epsilon\to 0}\int_{\mathbb{R}^{3}-B_{\epsilon}} fk^{i}k^{j}\partial_{i}\partial_{j}\phi \d V \ .
\end{eqnarray}
In the last equality the regularization of the integral at origin was used ($B_{\epsilon}$ is a ball of radius $\epsilon$). In our case we have $f=8\pi G\eta ^{2}+\frac{2GM}{r}$ and $\mathbf{k}=\d r - \d t$ when we use the equation (\ref{KS-form}) and leave out the prime sign on the time coordinate for simplicity. Here we can use the linearity of expressions (\ref{Ricci}, \ref{Scalar}) because the results for the second term in $f$ are already known from \cite{Balasin1}. The first term produces a regular distribution and can be calculated straightforwardly. Adding both together we get the following energy momentum tensor in the Kerr-Schild coordinates
\begin{eqnarray}
	8\pi GT^{a}_{b}&=&R^{a}_{b}-{\textstyle\frac{1}{2}}\delta^{a}_{b}R=\nonumber\\
	&=&-8\pi G \left(M\delta^{(3)}(x) + \frac{\eta^{2}}{r^{2}}\right) (\partial_{t})^{a} (dt)_{b}-\\
	&&-{\textstyle\frac{8\pi G\eta^{2}}{r^{4}}}(x\partial_{x}+y\partial_{y}+z\partial_{z})^{a}(xdx+ydy+zdz)_{b} \nonumber
\end{eqnarray}
Now we write the above result in a general Lorentz frame with momentum of the global monopole denoted by $P^{a}$ (where $P^{a}P_{a}=-M^{2}$) and the spatial direction in which we perform the boost by $Q^{a}$ ($Q^{a}Q_{a}=M^{2}$). Using the following expressions in terms of the Lorentz invariants 
\begin{eqnarray}\label{invariants}
M^{2}r^{2}&=&M^{2}\mathbf{x}\cdot\mathbf{x}+(\mathbf{P}\cdot\mathbf{x})^{2}\nonumber\\
\delta^{(3)}(x)&=&M\delta(\mathbf{Q}\cdot\mathbf{x})\delta^{(2)}({x}^{T})\\ 
M^{2}\vec{x}\d \vec{x}&=&M^{2}\mathbf{x}\cdot\d\mathbf{x}+(\mathbf{P}\cdot\mathbf{x})(\mathbf{P}\cdot\d\mathbf{x})\nonumber
\end{eqnarray}
and the corresponding relation for vectors (dot means a scalar product via $\eta$, bold letters stand for four-vectors and $x^{T}$ spans the spatial coordinate plane orthogonal to $\mathbf{Q}$) we can perform an ultrarelativistic boost (in which both $\mathbf{P}$ and $\mathbf{Q}$ turn into the null vector $\mathbf{p}$). We obtain the limiting energy momentum tensor (with the distributional term coming from \cite{Balasin1})
\begin{equation}\label{Final-EMT}
T^{a}_{b}=\left[\delta(\mathbf{p}\cdot\mathbf{x})\delta^{(2)}({x}^{T})+\frac{\eta^{2}}{(\mathbf{p}\cdot\mathbf{x})^{2}}\right]p^{a}p_{b}-\frac{\eta^{2}}{(\mathbf{p}\cdot\mathbf{x})^{2}}(\mathbf{p}\cdot\mathbf{\partial_{x}})^{a}(\mathbf{p}\cdot\d\mathbf{x})_{b}
\end{equation}
Evidently, this expression has, apart from delta distribution, singular behavior on the hypersurface $\mathbf{p}\cdot\mathbf{x}=0$. When one carefully considers the relation (\ref{invariants}) for the expression $M^{2}r^{2}$ one immediately concludes that the asymptotic behavior towards the limiting form is governed by $M^{2}\sim(1-v^{2})$ (appearing in the term $M^{2}\mathbf{x}\cdot\mathbf{x}$). It means that rescaling the solid angle according to the relation (\ref{angle-scaling}) cures the singular behavior, thus confirming the result of the previous section. On the other hand, the scaling considered in \cite{Lousto} is not sufficient to remove the singular behavior of the energy momentum tensor on the whole null hypersurface $\mathbf{p}\cdot\mathbf{x}=0$.

Finally, one could recover the result for the metric by solving Einstein equations for (\ref{Final-EMT}). When only the delta function term remains this reduces to solving a two-dimensional Poisson equation with source proportional to $\delta^{(2)}({x}^{T})$ which leads to the logarithmic solution of \cite{Aichelburg}. 

\section{Janis--Newman--Winicour spacetime}
Now, we would like to understand if the behaviour noted at the end of Section 4 (effective disappearance of the scalar fields and their influence in the boosting process) is natural for scalar fields at least in a highly symmetric situations. As a simple candidate for investigation we use a spherically symmetric, static and asymptotically flat solution of Einstein's field equations minimally coupled to a massless scalar field which can be described by the following line element \cite{JNW}
\begin{equation}\label{orig-metric}
ds^{2}=-f\left( R\right) dt^{2}+\frac{1}{f\left( R\right) }\left \{{dR^{2}}+\frac {1}{4}[(2R+r_{0})^{2}-\mu^{2}r_{0}^{2}]{d\Omega^{2}}\right \},\nonumber{}
\end{equation}
\[ {d\Omega^{2}}=d\theta ^{2}+\sin ^{2}\theta d\phi ^{2} \]
with
\begin{equation}
\ f(R)=\left [\frac {2R-r_{0}(\mu-1)}{2R+r_{0}(\mu+1)}\right]^{\frac{1}{\mu}}\ ,\ \phi={A}\ln f(R)
\end{equation}
and %
\begin{equation}\label{mu} 
\mu \equiv \sqrt{1+4A^{2}/r_{0}^{2}}\geq 1.
\end{equation}
In which A and $r_{0}=2m$ are two parameters. 

Our aim now is to find a form of the metric suitable for boosting, namely, the isotropic form. First, by choosing the following transformation we will bring the metric functions into a simpler form
\begin{equation}\label{transform1}
\tilde{R}=R+m\, ,\ M=m\mu{}
\end{equation}
Then the line element becomes
\begin{equation}\label{metric2}
\ ds^{2}=-f(\tilde{R}) dt^{2}+\frac{1}{f( \tilde{R}) }\left \{{d\tilde{R}^{2}}+(\tilde{R}^{2}-M^{2}){d\Omega^{2}}\right \},
\end{equation}
in which 
\begin{equation}\label{function2}
f(\tilde{R})=\left [\frac {\tilde{R}-M}{\tilde{R}+M}\right]^{\frac{1}{\mu}}.
\end{equation}
To get a conformally flat spatial part of the metric we apply one more change of variables
\begin{equation}\label{transform2}
\tilde{R}=\left({r}+\frac{M^2}{4r}\right)
\end{equation}
to finally get the desired form
\begin{equation}\label{final-metric}
\ ds^{2}=[g(r)-f(r)] dt^{2}+g(r)\left\{-dt^{2}+{dr^{2}}+{r}^{2}{d\Omega^{2}}\right\},
\end{equation}
\[f(r)=\left [\frac {1-M/2r}{1+M/2r}\right]^{\frac{2}{\mu}}\, ,\ g(r)=\left(1-\frac{M^{2}}{4r^{2}}\right)^{2} \frac{1}{f(r)}\ .\]
This metric has a singularity at $r=M/2$ as one can check from Kretschmann scalar.

We can try to obtain the Kerr-Schild form of metric in this case as well. For the line element (\ref{final-metric}) we would need to introduce a new radial coordinate $\tilde{r}=\sqrt{g(r)}r$. Since the boosting would then be performed in this new coordinate one can analyze the behaviour of the geometry only if one knows the explicit dependence of the metric functions on this new coordinate. However, inverting the relation $\tilde{r}=\sqrt{g(r)}r$ is not possible (for a general $\mu$) and although one might still attempt to draw conclusions based on an implicit relation the process would be extremely cumbersome and nontransparent. Another motivation to use the line element (\ref{final-metric}) stems from the possibility to exactly follow the steps of the original work \cite{Aichelburg}. 

The scalar field acting as a source of this geometry is given by the following prescription in the coordinates of (\ref{final-metric})
\begin{equation}\label{scalar-field}
\phi={A}\ln f(r).
\end{equation}
By putting $\mu=1$ and $A=0$, we obtain the Schwarzschild solution in isotropic coordinates. We will rewrite (\ref{final-metric}) in the corresponding cartesian coordinates
\begin{equation}\label{cartesian}
\ ds^{2}=g(r)\left \{-dt^{2}+dx^{2}+dy^{2}+dz^{2}\right\}+[g(r)-f(r)] dt^{2}
\end{equation}
where $r^{2}=x^{2}+y^{2}+z^{2}$. So the metric is separated into a conformally Minkowski part and one additional term which (together with the conformal factor) encodes the influence of the scalar field.

\section{Boosting JNW solution}

Now we apply the Lorentz transformation (\ref{Lorentz}) with $t=t'$ to the JNW solution described by the line element (\ref{cartesian}). Here, we again consider the scaling of the mass parameter (\ref{scaling}). So the line element (\ref{cartesian}) is in the following form after boosting
\begin{equation}\label{preboost}
\ ds^{2}=g(\bar {r})ds_{Mink}^{2}+\zeta(\bar{r}) \frac{(d\bar{t}-v d\bar{x})^{2}}{1-v^{2}}
\end{equation}
where
 \[ \zeta(\bar{r})=g(\bar {r})-f(\bar {r})\]
The Minkowski part in the line element (\ref{cartesian}) is obviously invariant under the Lorentz transformation. So the nontrivial computation consists of providing the boosting limits of the metric functions $g$ and $\zeta$. First, we check the limiting behaviour of the function $g(\bar {r})$
\begin{equation}\label{glimit}
\lim_{v \to \ 1} g(\bar {r})=\lim_{v \to \ 1} \left(1-\frac{M^{2}}{4\bar {r}^{2}}\right)^{2}\left [\frac {1+M/2\bar {r}}{1-M/2\bar {r}}\right]^{\frac{2}{\mu}} =\begin{cases} 1 & \bar{x} \neq \bar{t}\\{}\\ 1 &  \bar{x} = \bar{t} \end{cases}
\end{equation}
This means that the conformally Minkowski part becomes exactly Minkowski everywhere in the limit. Now, we study the limit of the crucial part, $\zeta(\bar{r})$,
\begin{eqnarray}\label{zetalimit}
\lim_{v \to \ 1} \zeta(\bar{r})=\lim_{v \to \ 1} &&\left\{\left(1-\frac{M^{2}}{4\bar {r}^{2}}\right)^{2}\left [\frac {1+M/2\bar {r}}{1-M/2\bar {r}}\right]^{\frac{2}{\mu}}-\right.\nonumber\\
&&-\left.\left[\frac {1-M/2\bar {r}}{1+M/2\bar {r}}\right]^{\frac{2}{\mu}}\right\}=\begin{cases}0 & \bar{x} \neq \bar{t}\\{}\\ 0 &  \bar{x} = \bar{t}\end{cases}
\end{eqnarray}
Considering the additional $\frac{1}{1-v^{2}}$ factor in the line element (\ref{preboost}) the limit is undefined. Expanding $\zeta(\bar{r})$ in $\frac{M}{\bar{r}}$ (which approaches zero for (\ref{scaling})) we get the following approximation 
\begin{equation}
\zeta(\bar{r})\simeq \frac {4M}{\mu \bar{r}}
\end{equation}
if we use the definition (\ref{transform1}) for $M$, we obtain exactly the same result as in the Schwarzschild case which is moreover independent of $\mu$. So the limit of the whole last term in line element (\ref{preboost}) turns out to be 
\begin{equation}
\lim_{v \to \ 1} \zeta(\bar{r}) \frac{(d\bar{t}-v d\bar{x})^{2}}{1-v^{2}}= 
\lim_{v \to \ 1} \frac {4p\sqrt{1-v^{2}}}{\bar{r}} \frac{(d\bar{t}-v d\bar{x})^{2}}{1-v^{2}}\nonumber
\end{equation} 
\begin{equation}\qquad\qquad=
\begin{cases}
\frac{4p}{|\bar{x}-\bar{t}|}(d\bar{t}- d\bar{x})^{2} & \bar{x} \neq \bar{t}\\
\\
\lim_{v \to \ 1}  \frac{4p}{\rho ({1-v^{2}})}(d\bar{t}- d\bar{x})^{2} &  \bar{x} = \bar{t}
\end{cases}
\end{equation}
where ${\rho}^2=(y^{2}+z^{2})$. For the case $\bar{x} \neq \bar{t}$ we obtain a flat spacetime which can be checked by computing the Riemann tensor. For $\bar{x} = \bar{t}$ either a singular transformation is needed \cite{Aichelburg} followed by a limiting procedure to obtain the well-known planar shock wave with the profile $\ln(\rho)\,\delta(\bar{x} - \bar{t})$. Or alternatively, one performs a distributional limit immediately and regularizes the result subsequently (in the manner of renormalization techniques of the Quantum Field Theory). Either way one ends up with the well-known shock wave which was originally obtained for Schwarzschild solution upon ultrarelativistic boost \cite{Aichelburg}. 

Since $f(\bar{r})\to 1$ in the limit (as seen by combining (\ref{glimit}) and (\ref{zetalimit})) and using (\ref{scalar-field}) the scalar field vanishes after boosting (everywhere except at the singularity $r=M/2$  of the metric) and because $\phi_{,r}\to 0$ its energy momentum tensor vanishes as well confirming the consistency of the above limit of the metric. So if one would like to apply the alternative boosting strategy \cite{Balasin1} where the left-hand side of the Einstein equations (considered as a definition of the energy momentum tensor) is boosted one immediately concludes that the regular part (corresponding to scalar field contribution outside of the singularity) does not contribute. However, we still have the singularity at the horizon position which gives a distributional contribution as in the Schwarzschild case \cite{Balasin1}. 

To make this qualitative statement more precise we can compute the distributional Einstein tensor for the line element (\ref{final-metric}). We do not have a Kerr-Schild form of the metric but we can use procedure for Schwarzschild metric in standard coordinates described in \cite{Balasin3}. This procedure is based on a regularization of the metric functions by introducing an arbitrary function (dependent on a parameter) that vanishes at singular point with subsequent removal of the regularization function influence on the Einstein tensor via limit in the parameter. In the case of our metric (\ref{final-metric}) we can introduce the following regularized metric functions (inspired by \cite{Balasin3}) parametrized by $\lambda > 0$
\begin{equation}
	f(r)=\left [\frac {1-M/2r}{1+M/2r}\right]^{\frac{2}{\mu}+\lambda},\ g(r)=\left(1-\frac{M^{2}}{4r^{2}}\right)^{2-\lambda} \frac{1}{f(r)}.
\end{equation}
As proved in \cite{Balasin3} the regularization result does not depend on the specific regularization function provided it is smooth and vanishes at the singular point. We will show the explicit computation in the case of Ricci scalar since the nontrivial components of the Ricci tensor (in mixed components form) have the same behaviour near the singularity $r=M/2$
\begin{equation}\label{Ricci-regularized}
	R=\frac{32 M^{2}}{4^{\lambda}\mu^{2}} \frac{(2r-M)^{2\lambda+\frac{2}{\mu}-4}}{(2r+M)^{\frac{2}{\mu}+4}} r^{2-2\lambda} K(r)\ ,
\end{equation}
where $K(r)$ is a regular and nonzero quadratic function in the vicinity of the singularity $r=M/2$. Now, we concentrate on the singular part of (\ref{Ricci-regularized}) and consider it as a distribution on the space of smooth test functions with compact support. As in the case of the global monopole we investigate just the spatial distributions and due to a spherical symmetry we are only concerned with the radial dependence of our test functions. We note that the two-dimensional spheres at $r=M/2$ have vanishing area so it is useful (and geometrically reasonable) to simplify calculations by introducing the shifted radial coordinate $\rho=r-M/2$ and having the test functions $\phi(\rho) \in C^{\infty}_{0}(0,\infty)$. The singular part of (\ref{Ricci-regularized}) when considered as a distribution can be decomposed in the following way \cite{Balasin3}
\begin{eqnarray}
	\left(\rho^{2\lambda+\frac{2}{\mu}-4},\phi\right)&=&4\pi\int_{0}^{\infty} \rho^{2\lambda+\frac{2}{\mu}-2} \left[{\phi}-{\phi}(0)\Theta(1-\rho)\right] d\rho+\frac{4\pi{\phi}(0)}{2\lambda+\frac{2}{\mu}-1}=\nonumber\\
	&=&\left(\left[\rho^{2\lambda+\frac{2}{\mu}-4}\right],\phi\right) + \frac{4\pi}{2\lambda+\frac{2}{\mu}-1}(\delta,\phi)
\end{eqnarray}
where we have divided the integration into a regular part (where the test function is shifted to vanish at $\rho=0$) and a singular part which gives rise to $\delta$-function. The specific regularization chosen above is valid for $2\lambda+\frac{2}{\mu}-1>0$ which implies $\mu<2$ upon taking the limit $\lambda \to 0$. Applying the procedure to the whole Einstein tensor one recovers the singular part identical to the Schwarzschild solution and the regular part coming from the scalar field presence. Upon boosting the regular part trivially vanishes and the delta function generates a standard Aichelburg-Sexl shock wave when one solves back for the metric that is giving rise to such a boosted energy momentum tensor (as is the case for the global monopole or more explicitly in \cite{Balasin1}). So the above result is confirmed even when using the second approach to boosting. 

Note that if we want to keep $\mu=const.$ in the limiting process we have to scale parameter $A$ like the mass because of (\ref{mu}) and (\ref{transform1}).

Finally, let us add that for the original line element (\ref{orig-metric}) the range of coordinate $R$ is $(\frac{1}{2}r_{0}(\mu-1),\infty)$ which translates into $r\in (\frac{M}{2},\infty)$ for the line element (\ref{final-metric}) used for boosting. Due to the scaling of the mass (\ref{scaling}) the range of both $r$ and $\bar{r}$ becomes $(0,\infty)$ making the Schwarzschild limit correct even from this point of view.

\section{Conclusion}
We have derived a consistent ultrarelativistic limit of a general global monopole solution (with the monopole charge sourced either by a triplet of scalar fields or by a cloud of strings) using two approaches. The main result is the necessary rescaling of the deficit solid angle in order to obtain a well defined solution resulting from boosting the original static metric. Both methods give the same scaling relation. We have also commented on a previous result \cite{Lousto} suggesting a different scaling of the solid deficit angle and argued what are the benefits of considering the above derived prescription and its consistency with the behavior of the source fields. Additionally, we have analysed the scaling with respect to the behavior of the spatial volume element which gave us a possible interpretation for the constant $\eta_{0}$ appearing in the scaling relation.

From the physical point of view the vanishing of the solid deficit angle might be considered natural for an observer moving infinitely close to the speed of light due to the relativistic contraction which effectively squeezes the transversal directions thus making the solid deficit angle irrelevant.

Since the triplet of scalar fields acting as a source of the global monopole effectively vanishes in the boosting limit we have tried to investigate whether this suggests a general behavior of scalar fields in highly symmetric situations. As a good candidate we selected the well-known spherically symmetric solution --- the Janis--Newman--Winicour spacetime. The ultraboost limit of this geometry approaches the Schwarzschild solution (which is at the same time the vacuum limit as shown in \cite{JNW}) with no trace of the scalar field left. This result is confirmed by checking the behaviour of the scalar field itself or applying the alternative approach to boost. Interestingly, the ultrarelativistic boost washes away any trace of the scalar field and always recovers the Aichelburg--Sexl geometry irrespective of the field strength. This behavior is analogous to the one noted in \cite{Bicak} for the ultrarelativistic boost of a curved spacetime coupled to an electromagnetic field, namely that the necessary rescaling of the charge leads to the weak-field regime which is insensitive to the potential nonlinear dynamics of the electromagnetic field (while the boost in the flat case is influenced by a specific nonlinear theory of electromagnetic field). However, one should confirm this observation by really calculating the ultrarelativistic boost of such models.

So it seems that scalar-field sources in static, highly symmetric geometries tend to vanish in the ultraboost limit.

\section*{Acknowledgments}
This work was supported by grant GA\v{C}R 14-37086G. We thank M. \v{Z}ofka for the language corrections of this manuscript.

\end{document}